# VALIDATION OF THE DEVELOPMENT METHODOLOGIES

Ammar LAHLOUHI

Department of computer science, University of Batna, 05 000 Batna

## ABSTRACT

*This paper argues that modelling the development methodologies can improve the multi-agents systems software engineering. Such modelling allows applying methods, techniques and practices used in the software development to the methodologies themselves. The paper discusses then the advantages of the modelling of development methodologies. It describes a model of development methodologies, uses such a model to develop a system of their partial validation, and applies such a system to multi-agent methodologies. Several benefits can be gained from such modelling, such as the improvement of the works on the development, evaluation and comparison of multi-agent development methodologies.*

## 1. INTRODUCTION AND MOTIVATION

The software engineering aims the study of rigorous scientific approaches that allow reducing the costs of the software development and improving its quality. The enhancements of the formalisms (mathematical formalisms, graphical notations…) have largely contributed in reducing the costs and in improving the quality. However, with the experiments and failures in the software development, it was noted that these enhancements are insufficient and we realized the importance of improving the software development process (software process, in short). The software process was invented by Lehman in 1969 (Programming process [21]) after the software crisis of the sixties. It regards the development of software as a process similar to those used in other industries. At the beginning, the (classical) software process is viewed as unique since, at this time, the practices use functional models exclusively. The works on the software process then targeted the totality (include all the development phases) and the unification (one process for all systems models).

The use, thereafter, of the classical software process to the development of object-oriented systems (respectively, knowledge based systems (SBK)) showed, very quickly, the inadequacy of such process. The adaptation of this latter to object-oriented modelling (respectively, SBK) has lead to object-oriented software engineering (respectively, SBK engineering). Similar preoccupations appeared with the multi-agent modelling and the agent-oriented software engineering (AOSE) were born and several works were undertaken to establish its bases (e.g. [19]). A consensus was then established on the fact that the software process of multi-agent development is different from those of other models.

The development of software systems includes primarily the phases of requirements analysis, design and implementation. The aim of the analysis phase results is not to be refined so that they will be carried out by machines, but rather they are intended to be used by the humans for their comprehension and their reasoning. If we improve it, such results will be valid for all the models as long as they do not comprise conceptual aspects of a particular model. The analysis phase can then be improved in parallel to the models' evolution and in an independent way. This is not the case of the design phase. A new model (object-oriented, multi-agent…) introduces new abstractions which do not concern the requirements phase in a direct way. The difference between the software processes of the various models lies then mainly to the design phase. When a new paradigm is adopted, the works on its development process targets particularly the design phase since the novelties will be introduced particularly in such phase.

The implementation phase depends closely on the design. It must then include directives allowing the implementation of the conceptual models (design phase's results). The phases of design and implementation constitute together the system's development. The process of multi-agents systems development is particularly specific to its development phase.

The description of the development process is given in the development methodologies which include the essence of what a new model provides compared to the more traditional models. Considerable efforts were invested by the AOSE community to equip the multi-agent paradigm with good methodologies. Significant advancements were then recorded especially in organisational methodologies. Several methodologies proclaiming to be multi-agent were pro-posed (see [18, 35, 38], for a survey) so much that their number already exceeds those of the more traditional models methodologies. In spite of such efforts, the multi-agents methodologies did not achieve the goals pursued by the AOSE community:

(1) they aren't adopted largely in the software industry,
(2) little of them continue to be developed and to be used,
(3) some were abandoned in spite of the fact that their preliminary results appeared promising.

Consequently, improvements were made to interesting methodologies (MaSE [11], Gaia [41], tropos [2]. . . ). In addition, some platforms were proposed to evaluate methodologies [6, 7, 9, 33, 34] and others to compare them [10, 25, 32] in order to identify their insufficiencies. The works undertaken in this field and the approaches suggested are significant and they are focused on the essential question which is that of the methodologies. In spite of these efforts, multi-agents methodologies did not succeed to surpass their current drawbacks.

The majority of the multi-agents methodologies difficulties are not specific to them. Traditional methodologies (especially object-oriented and engineering SBK) and the software processes also knew and continue to experience difficulties. However, works on the software process have advanced where they succeeded in producing development environments, standards (such as CMM [30], CMMi, …), etc. which improved the software development.

Historically, the advancement is the result, particularly, of the turning point which the software processes knew in the middle of the Eighties. This is the case especially with the famous paper of Osterweil (entitled "software processes are software too" [27]) that he claimed it in 1997 (ten years later) as a position paper ("software processes are software too revisited" [28]). Osterweil stipulates that the software process is also a software system (a domain of study) to which one can apply methods, techniques and practices used usually in the software development. Several researchers proposed then models for the software process aiming various objectives: realization of software development environments (such as [31]), improvement of the software process (such as [32]), and simulation of the software process targeting its analysis (such as [20]).

This paper stands a thesis stipulating that a solution of the difficulties of the multi-agents methodologies can be sought, partly, in their consideration as a domain of study in a way similar to that followed in the software processes. Development methodologies of particular models will be then treated as particular applications. For the moment, it is certainly hard to concretize this step in a rigorous way. However, as we will show it in this paper, even a partial concretization is beneficial. On the other hand we are persuaded that, at the long term, its concretization is possible and, moreover, it opens many prospects.

The methodologies modelling will benefit from the advantages of the modelling where it will allow, among other things:

1. Making easier the reasoning in analysing and creating methodologies, where such a model shows the main relevant aspects (to be not ignored) and only them (to not consider other aspects erroneously or unnecessarily),
2. Simplify the exchange and the discussion of the methodological knowledge,
3. Guide the "development" of the development environments that allows developing systems following such methodologies. This facilitates the experimentation (so important) of the methodologies and their analysis,
4. Directs the development of new methodologies where it identifies the required elements (which a methodology's developer must seek) that will be made explicit and clear.

For a first concretization of the previous thesis, this paper proposes a model of methodologies, uses it to the development of a system of partial validation of methodologies and then applies this system to multi-agents methodologies. The paper shows the feasibility of such thesis while emphasizing its benefit even if it is partial.

The remainder of this paper is organized in four sections. Section 2 describes the model of methodologies. Section 3 describes the system of their partial validation. Section 4 describes the use of this system to the partial validation of multi-agents methodologies. Section 5 discusses the contributions of the paper and concludes it.

## 2. MODELLING THE SYSTEMS DEVELOPMENT METHODOLOGIES

Modelling is the activity allowing the development of models. It depends on what we want to do with the model. This can be:

1. Development: The model will be used as means of design's description in order to carry out a new reality. This last can replace the existing one if it is regarded as being better than the existing one,
2. Simulation: The model will be used for the development of a simplified reality of complex one. The simplified reality does include only essential elements useful for reaching the goal of the simulation in question. The processing (manual or automatic) of the simplified reality gives results which one can analyze to draw some conclusions,
3. Analyze: The model will be useful like a simplified representation of a complex reality to facilitate processing which is difficult to be carried out directly on such reality.

Our objective, in this section, is to develop a model of methodologies which one can use as means facilitating the reasoning on the methodologies. Consequently, this model will contain only pertinent aspects of the methodologies. It can be used to any aim, such as validation, foundation or improvement of methodologies.

This section is organized in three subsections. Subsection 2.1 recalls some basic concepts which will clarify the terminology used in the presentation of the model. Subsection 2.2 describes the proposed model.

**2.1 Basic Concepts**

A model is an abstraction of some reality where it comprises some elements of this reality considered as being relevant and ignores others. For example, a computer science models considers only informational aspects of realities and ignores all other aspects (physical, biological ...).

A meta-model makes it possible to limit the relevant elements being able to be represented in a class of models. For examples, the following traditional meta-models are well-known in computer science:

1. Analytical (functional or processing based): It allows limiting its models to include only data processing,
2. Systemic (or data based): It makes it possible to consider, in addition to processing, data on the realities' states (it is less abstract than the analytical model).

To simplify the models description, we use formalisms. The latter propose some notations to be used in describing the models. These notations are abstractions which are in direct relation with those defined in the meta-model. Usually, the formalism and the meta-model are confused. However, they are different:

1. The meta-model is validated compared to a reality, i.e., according to whether it reflects targeted reality or not,
2. The formalism is validated compared to the meta-model, i.e., according to whether it reflects the meta-model that it represents or not.

Depending on the targeted objective, formalisms can be mathematical (based on a formal semantics allowing the demonstration of some properties), composed of convivial graphical symbols (simplifying models description), mixed of these two forms, etc.

The development of a model is done following a modelling process whereas its development is done according to an implementation process. The first one is a prescription of a procedure (combination of stages) for identifying the relevant elements of the reality which are defined in the meta-model. On the other hand, an implementation process makes it possible to identify the reality's elements, which correspond to the abstractions described in the models that allow creating a new reality that reproduces the model. The two processes are then based on an association of elements of the reality (existing or imagined) to abstract elements of the meta-model.

Table 1. A first systemic model of methodologies.

| **Data model** | | **Processing model** |
|---|---|---|
| Meta-model | Formalism | Methodological process |

Table 2. A concept description.

| Identifier | Significance |
|---|---|
| Processing | A processing is a composition of actions to undertake on data in order to obtain targeted results |

Table 3: A notation description with a graphical symbol.

| Symbol | Semantics |
|---|---|
| 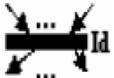 | Transition representing a processing of identifier Id, whose control flow including synchronization is expressed by tokens |

Table 4: Data model of systemic modelling of methodologies.

| Methodological element | Necessary elements |
|---|---|
| Methodology | <Meta-model, methodological Process> |
| Meta-model (MM) | {<Identifier, Significance: Necessary elements: { ... }, Preferable elements: { . . . }>} |
| Formalism (F) | {<Symbol, Semantics: Necessary elements: { ... }, Preferable elements: { ... }>} |
| Methodological Process (MP) | <Combination, {Stage: Necessary elements: { ... }, Preferable elements: { ... } >}}> |
| Relation F — MM | {Relation type:{< <Notation, Semantics element>, <Concept, Significance element> > }} |
| Relation MP — MM | {Relation type: {< <Stage, Stage element>, <Concept, Significance element> > } } |
| Relation MP — F | {Relation type: {< <Stage, Stage element>, <Notation, semantics element> >} } |

## 2.2 Model of Methodologies

An analytical modelling of methodologies will be reduced to a composition description of the development stages (processing to be carried out). Such modelling does not take into account the models handled by the processing. A systemic modelling (see Table 1) is then preferable to it. This last one includes a data model on which the processing will be carried out. In this case, the composition of stages forms what we call the methodological process. The data model to be handled is that of "the models". It is the associated formalism or the meta-model. A methodology will be then defined as a couple <Meta-model or Formalism, Methodological process>.

This paper is limited to a systemic modelling. Such modelling is sufficient for reach our objective of this paper, which is that of partial validation of methodologies. It isn't necessary to seek object-oriented or multi-agent modelling. Moreover, systemic models are simple to present, and well known, used and formalized than object-oriented and multi-agent modelling. The following paragraphs presents in more details the data and processing models of methodologies.

### 2.2.1 Data

The meta-model is a set of concepts which will be used to limit the relevant aspects of the reality to be represented in models. It is a set of couples <Identifier, Signification> (see table 2, for an example of concept description). The identifiers are elements used to express models as well as the semantics of the formalism's notations. Significance is a description making it possible to specify the meaning which one assigns to a concept in the reality.

If the formalism is present, its description depends on the meta-model. Otherwise, the data model will be reduced to the meta-model only. A model of methodologies can then be described as in table 1. The following paragraphs describe the data model and how one can use it to develop processing around it.

Formalism is a set of notations where each one is associated to a meta-model's concept and is described like a couple <Symbol, Semantics> (see table 3, for an example of a notation description). A symbol is a textual or graphical element which will be used in models descriptions. Semantics is the meaning associated to a symbol. Its description is made in accordance with the concept's significance to which the notation is associated.

In table 4, we used the following notations:

- braces { and } delimit sets elements like their usual use in mathematics,
- brackets < and > delimit the tuples,

The preferable elements have the same description as the necessary elements except in the case of methodology arrow where we have the "formalism" as a preferable element. In the composition of the necessary elements and the preferable elements, one can also find elements which are necessary and those which are preferable.

A methodological process is a combination of stages where its implementation will result, starting from the requirements, in a development of a new reality. A stage allows a description (design) or an implementation of one or several elements of a model.

A methodology must contain elements which, without them, such methodology will be considered as wrong. It must also contain other elements that allow simplifying the development. The presence of these latter elements is important. Consequently, we subdivide the elements of a methodology in two classes: Necessary elements and preferable elements.

We summarize the various elements of the model of methodologies explained, until now, in the form of a data model given in table 4. Such a model can be described using one of existing formalisms such as a semantic model. Moreover, more other aspects can be added to this model, such as the details of significances, of semantics and of stages, integrity constraints, etc. However, we will not do it here because we cannot completely formalize some aspects relating to the methodologies, especially in general form, such as the description of semantics model. On other hand, the description of table 4 will be sufficient to achieve our first goal which is that of showing the benefits of methodologies modelling.

**2.2.2 Processing**

Processing can be information seeking on the model's elements, updating data or particular applications such as partial validation of methodologies. For examples, the updates can be subdivided in three processing:

1. Addition of (build) a new methodology,
2. Suppression (discard) of an existing methodology,
3. modification (revision) of an existing methodology.

For example, building a new methodology according to the methodologies data model (given in table 4) can be made as the following process:

1. Describe a meta-model,
2. Propose a formalism,
3. Build a methodological process that can be done as follows:
    - Propose a set of stages,
    - Propose a combination of such stages:
        o Organization of the stages in: top down, bottom-up or mixed process,
        o Succession of the stages: sequence, parallel, synchronization ... ),
    - Detail each stage.

Several other processing can be developed around the data model of methodologies such as the example of methodologies validation which we describe in section 3.

## 3. PARTIAL VALIDATION OF METHODOLOGIES

In spite of its strong abstraction, the proposed model of methodologies is already informative. A good exploitation of such information can be of a great help in the foundation of methodologies, their validation and their evaluation. For examples, it indicates that:

1. a good methodology must include also a meta-model and formalism,
2. it shows the need for relations between the formalism and the meta-model, and between the methodological process and the formalism or the meta-model.

Since the methodologies contain very creative tasks which cannot be formalized, it isn't possible to handle its aspects entirety. The description of a methodologies model can contain only the elements making it possible to achieve a given goal. This section proposes then an application of the methodologies model (described in section 2) to build a system of partial validation which can be considered as a verification of minimal required aspects. The objective of this system is to check if a methodology is valid. It isn't concerned with its evaluation, ease of use, comparing it, etc.

The remainder of this section is organized in four subsections. Subsection 3.1 gives a characterization of the methodologies validation. Subsection 3.2 describes a system of partial validation of methodologies.

**3.1 Characterization of partial validation of methodologies**

A methodology targets certain aspects of the reality. For instance, a functional methodology targets data processing and a robotics one targets robotics systems. A meta-model is correct if the association of its concepts with the targeted reality is correct. Let us note that, at the moment, such association cannot be completely objective. The role of a methodology is to create and implement models which descriptions are based on the concepts of a correct meta-model. A methodology is correct if it takes into account the all models that it targets and only these one. Its validity can be then measured by the models it allows to describe and achieve effectively. This measurement can be made in relation to the targeted models. We say that a methodology is valid if it comprises the all elements where their exploitation allows modelling and/or developing targeted realities and only them. In this subsection, we try to characterize the targeted models. We give four types of characterizations that we can use in the methodologies validation.

We regard the elements of the methodology M allowing the development of targeted realities, and only those realities, as necessary elements. Their absence brings to confusions. In this case, the set of the all models, that such M allows to develop, will include models that aren't part of those targeted by M. This is a bad generalization of the methodology, qualified of excessive generalization (EG).

At the opposite of necessary elements, the presence of some other elements makes it possible to develop:

1. only realities that not belong to targeted one, or
2. do not allow implementing any reality.

We qualify them of erroneous elements. This case is that of very bad specialization and we qualify the presence of such elements of abusive specialization (AS).

The models are abstractions where their realities comprise additional elements which cannot be common to all concerned realities. If a methodology requires the presence of some of these additional elements, it reduces the number of targeted realities. These elements are qualified of superfluities. Their presence in a methodology makes it possible to build only models that contain these superfluities elements and draws aside, wrongfully, those which do not contain them. This case is that of bad specialization which qualified of excessive specialization (ES).

The main objective of a methodology is to develop models and to implement them correctly, comes, then, the simplification of modelling and development. The presence of some elements in a methodology targets the development simplification. Their presence is not necessary but rather preferable. They are qualified of preferable elements. Their absence does not bring to errors but prevent the methodology from benefiting from simplifications that they offer. Their absence is qualified of undesirable generalization (UG).

Checking the partial validity of methodologies will concern checking the absence of the necessary and preferable elements and the presence of the superfluities and erroneous elements. Whereas checking these characterizations is difficult and, sometimes, can bring to subjective judgments, they are of primary importance especially to conduct discussions on methodologies aspects. The subsection 3.2 shows how these characterizations can be used in the partial validation of the methodologies.

1. Table 5: Processing associated to the validation of methodologies

| Processing | Inputs | Outputs |
|---|---|---|
| **Create necessary elements set** | According to its implementation | Necessary elements |
| **Create preferable elements set** | According to its implementation | Preferable elements |
| **Identify necessary elements** | Set of methodologies necessary elements, Methodology | Methodology where necessary elements are marked, Set of necessary elements absent from the methodology |
| **Identify preferable elements** | Set of methodologies preferable elements, Methodology | Methodology where preferable elements are marked, Set of preferable elements absent from the methodology |
| **Check the presence of superfluities elements** | Methodology where necessary elements and preferable elements are marked | Methodology where superfluities elements are marked, Set of superfluities elements present in the methodology |
| **Note erroneous elements** | Methodology where necessary, preferable and superfluities elements are marked | Methodology where erroneous elements are marked, Set of erroneous elements present in the methodology |

## 3.2 System of Partial Validation of Methodologies

According to the methodologies model, three lines were defined where the characterizations (EG, AS, ES and UG, described in the subsection 3.1) will be applied to them for the methodologies validation. These lines are:

1. Global line: The methodology's components (meta-model, formalism and methodological process) are handled in a global way,
2. Local line: Each methodology's component is handled in an isolated way,
3. Relational line: The relations between the components and between the constituents of these components are handled.

In the global line, we can already make the following remarks:

1. It is preferable that the methodology includes a formalism: its absence is then a UG,
2. It is necessary that the methodology comprises a meta-model and a methodological process: The absence of one of them is an EG,
3. The presence of elements, others than the three previous components, will be regarded as an AS.

In the local line, the application of the characterizations to the components and their constituents brings to the following remarks:

1. The absence of a necessary constituent (such as the absence of a necessary concept) is an EG,
2. The absence of a desirable constituent is a UG,
3. The presence of superfluities constituent is an ES,

    4. The presence of an erroneous constituent is an AS.

Depending on the details included in the model of the methodologies, the characterizations can be applied recursively to the components regarding them as being composed of properties. If the used model comprises these details, the method of validation can handle them. Let us note that other elements can be added to the proposed model in order to complete the validation. If such elements are added, the resulted model will be another model than that used in this system of validation.

In the relational line, the characterizations will be applied to the relations existing between components and between the constituents of the various components as follows:

1. The absence of necessary relations is an EG,
2. The absence of preferred relations is a UG,
3. The presence of superfluities relations is an EG,
4. The presence of erroneous relations is an AS.

For example, the semantics of a notation must be in conformity with the significance of the concept which it represents. The absence of this conformity relation is regarded as an EG. Where it is desirable that the conformity relation will be described formally description.

The process of partial validation can be summarized into two stages:

    3. The first stage (initialisation): it will prepare the useful aspects to apply them in the validation of the methodologies. This first stage starts with the choice of a methodologies model and then the creation of the sets of the necessary and preferable elements,
    4. The second stage: relates to the used approach to validate a particular methodology. It starts with the methodology's analysis in order to identify the methodological knowledge elements. Then, it identifies the necessary elements and the preferable elements present in the methodology. Lastly, it checks the presence of the superfluous elements and notes the remaining elements as erroneous. During these identifications, it is necessary to take into account synonyms and homonyms. The various procedures used by the process of the validation are described in table 5.

## 4. APPLICATION OF THE PARTIAL VALIDATION SYSTEM TO MULTI-AGENTS METHODOLOGIES

This subsection describes an application of the system of partial validation to some multi-agent methodologies. Its objective isn't to compare these methodologies, to criticize them objectively, to note advantages of some of them (which can be numerous), or …. It presents only the results of using the previous system to evaluate such methodologies without detailing their justifications. It presents these results according to the outputs of the system of partial validation.

We choose to validate some methodologies instead of others because we think knowing them better, they are more referenced in the literature, and their documentation is available online.

The partial validation must be based on a model of methodologies. However, this paper will use some elements of this model which have some consensus in the community of multi-agent methodologies to avoid the current discussions of their concepts (agent, cooperation...). Another reason for this choice is the simplification of the validation's description.

The remainder of this subsection is subdivided according to the previous validation lines of the system of partial validation: The global validation is described in 3.3.1, the local validation in 3.3.2 and the relational validation in 3.3.3.

## 4.1 Global Validations

The study of the literature on multi-agent methodologies shows that many of these methodologies do not consider the formalism. They present then a global UG. The meta-model and the methodological process are present in methodologies except in some atypical cases, such as the absence of the methodological process in [26] and the absence of the meta-model in [22] which present then a global EG.

## 4.2 Local Validations

In the following, only the methodological process is considered without the associated meta-model and its formalism. To validate the methodological processes of multi-agent methodologies, a model of methodological process of multi-agent systems is used. This process is based on the following observation: "The multi- agent systems must be considered from two complementary viewpoints, individual and collective. Moreover, a consensus starts to emerge, in the multi-agent systems community, on the fact that collective aspects are organizational."

The model of multi-agent methodological process used here comprises the three following stages:

1. Organization: Develop an organization model on the basis of the system's requirements,
2. Modelling: Derive a multi-agent model from the organization,
3. Implementation: Implement the multi-agent model as a multi-agent system.

The partial validation of the methodological process must take into account the stages and their combination. However, the combination isn't included here. The partial validation relates to the absence of necessary (EG) and preferable (UG) stages and the presence of superfluities (ES) and erroneous (AG) stages. The result of the validation of some methodologies brought us to the results described in table 6.

Table 6: Examples of surveyed methodologies.

| Missing stage | Methodologies examples |
|---|---|
| Organization (NDG) | UML's extension[24], MESSAGE/UML[4], MESSAGE[16], MAS-CommonCADS[12, 17], MaSE[36], Tropos[15], SODA[26], Prometheus[29], ADELFE[1], Zeus[23] |
| Modeling (EG) | UML's extension[24], MESSAGE[16], MESSAGE/UML[4] |
| Implementation (EG) | Gaia[37, 41], Alaaddin[13, 14], Cassiopeia[8], Styx[3] |

### 4.3 Relational Validations

In the following, we discuss the relation of formalisms with its meta-model. This discussion concerns the use of coloured Petri nets like a formalism for multi-agent systems modelling, as used in G-Net [39, 40], and its relation to the multi-agent meta-model. An ordinary Petri net comprises:

1. A set of transitions that represents the processing,
2. A set of arcs and a set of places binding these transitions between them and representing the control flow between the transitions with inscriptions of tokens on arcs and places allowing synchronization of transitions processing.

According to this description, an ordinary Petri net is then analytical concurrent. The coloured Petri nets consider, in addition, colours and colours functions which represent data. They are then systemic concurrent. They are insufficient to completely describe multi-agent models. Many concepts such as agent's control, communication by speech acts not by data flow only … do not have associated notations in coloured Petri nets. The absence of these notations makes several relations between the formalism of coloured Petri nets and the multi-agent meta-model missing (EG). Note that this observation does not relate to the hybrid formalisms using coloured Petri nets for modelling some aspects of the multi-agent systems supplemented by other notations.

## 5. CONCLUSION AND DISCUSSION

Some works can be regarded as being related to the modelling of the methodologies such as the project of FIPA methodology [5], evaluation of the methodologies [6, 7, 9, 33, 34] and their comparison [10, 25, 32]. The objective of the FIPA methodology is to develop a particular methodology by unifying various aspects of existing methodologies (methodologies fragments). This is an interesting work for the development of a new methodology by taking into account good aspects of existing methodologies while extending them when needed. The evaluation and comparison of methodologies consists in defining a set of criteria to evaluate methodologies in order to compare them. This allows developers to choose the best methodology adapted mainly to the specificities of their applications. Such works are significant where they allow not only a comparison of the methodologies but also to make decisions on their improvements. The previous works can be significant applications of the methodologies model. However, since they do not consider explicit modelling of the methodologies, they do not have the same scope. We think that the results of these works would be better if they are based on a methodologies model.

The major difficulty running up against the modelling of the methodologies are related to the development of a complete model. The exhaustive identification of the elements of such a model is not obvious and cannot have the necessary consensus for its acceptance and exploitation. However, we think that this modelling is the way allowing foundation, validation, evaluation and improvement of the methodologies in a more easier and rigorous manner. Moreover, the incomes of such modelling are numerous even with say a partial model (which is also a model). This is because any contribution in the modelling of the methodologies is a multiplication of the benefits of those that can be gained from the applications since it addresses not only one methodology but several of them. Note that the benefits of methodology are the sum of the benefits gained from the development of particular applications using this methodology.

The importance of the methodologies modelling in the simplification of the reasoning is showed in the identification of the validation elements. The difficulties of the methodologies validation (as described in this paper) are due to the difficulty to exhaustively determine the set of all the necessary elements and the preferable one. It is due also to the difficulty to verify objectively the presence of superfluities elements. The minimal set of necessary elements can be obtained by a compromise between the various methodologists. This will make it possible to perform a partial validation (since a partial of "partial validation" is also a "partial validation"). Whereas this set can be enriched as the understanding of the models and methodologies becomes better. However, the identification of the superfluities elements will remain very intuitive (not formal). The difficulties relate also to the analysis of a given methodology to identify the elements of methodological knowledge where this analysis must be made in a rigorous way. Whereas these difficulties are embarrassing, similar difficulties are also those of any handling methodologies such as their evaluation and comparison. Although the validation given in this paper as an application of the methodologies model, it presents some significant points such as the unification of the validation aspects in three lines (global, local and relational). This is made possible thanks to the methodologies model. The example shows then how the model can contribute in simplifying and justifying the reasoning on the methodologies by exploiting the relevant elements and only them.

The work presented in this paper opens, in the foreseeable future, various prospects concerning problems which are actually subject of the debate in the multi-agent systems community such as open and self-organized systems. Currently, and in the foreseeable future, our work focuses on the development of the environments for methodologies. This is beneficial since the methodologies necessitate rapid experimentations during their development and their improvement. We propose then to model the methodologies as a multi-agent model which can be used to develop such environments. This model will be open and self-organized so its adaptation to particular methodologies will be simple. It benefices then from the work described in this paper.

## REFERENCES

[1] Bernon, C., Gleizes, M. P., Picard, G., & Glize, P. (2002). The Adelfe Methodology For an Intranet System Design. In proceedings of the fourth international bi-conference workshop on agent-oriented information systems (AOIS 2002 at CaiSE'02) Toronto (Ontario, Canada), May 27-28, 2002.

[2] Bresciani, P., Perini, A., Giorgini, P., Giunchiglia, F., & Mylopoulos, J. (2004). Tropos: An agent-oriented software development methodology. Autonomous Agents and Multi-Agent Systems, 8(3), 203-236.

[3] Bush, G., Cranefield, S., & Purvis, M. (2001). The Styx agent methodology. The Information Science Discussion Paper Series 2001/02, Department of Infor- mation Science, University of Otago, New Zealand., Jan. 2001.

[4] Caire, G., Coulier, W., Garijo, F., at al.. (2002). Agent oriented analysis using MESSAGE/UML. In Agent-oriented software engineering II (pp. 119-135). Springer Berlin Heidelberg.

[5] Calisti M., M. Cossentino, Methodology Work Plan, Document number f- wp-00023, Approved at 2003/02/10, http://www.fipa.org/docs/wps/f-wp-00023/f-wp-00023.html (accessed 27/04/2014)

[6] Cernuzzi, L., & González, M. (2002). A Framework for Evaluating WIS Design Methodologies. Enterprise Information Systems III. Kluwer Academic Publishers, Dordrecht, The Netherlands, 198-207.

[7] Cernuzzi, L., & Rossi, G. (2002). On the evaluation of agent oriented modeling methods. In Proceedings of Agent Oriented Methodology Workshop, Seattle (Vol. 29).

[8] Collinot, A., Drogoul, A., & Benhamou, P. (1996, December). Agent oriented design of a soccer robot team. In Proceedings of the Second International Conference on Multi-Agent Systems (ICMAS-96) (pp. 41-47).

[9] Cuesta, P., Gómez, A., González, J. C., & Rodríguez, F. J. (2003). A framework for evaluation of agent oriented methodologies. In Proceedings of the Conference of the Spanish Association for Artificial Intelligence (Vol. 147, pp. 151-152).

[10] Dam, K. H., & Winikoff, M. (2004). Comparing agent-oriented methodologies. In Proc. of the Fifth Int. Bi-Conference Workshop on Agent-Oriented Information Systems (at AAMAS03), pp. 78-93. Springer Berlin Heidelberg.

[11] DeLoach, S. A., Wood, M. F., & Sparkman, C. H. (2001). Multiagent systems engineering. International Journal of Software Engineering and Knowledge Engineering, 11(03), 231-258.
[12] Elammari, M., & Lalonde, W. (1999). An agent-oriented methodology: High-level and intermediate models. In Proc. of the 1st Int. Workshop. on Agent-Oriented Information Systems (pp. 1-16).
[13] Ferber, J., & Gutknecht, O. (1998). A meta-model for the analysis and design of organizations in multi-agent systems. In Proceeding of the 3rd International Conference on Multi-Agent Systems (ICMAS 98). IEEE CS Press.
[14] Ferber, J., Treur, J., Müller, J. P., Gutknecht, O., & Jonker, C. M. (2000). Organization models and behavioral requirements specification for multi-agent systems. In Multi-Agent Systems, International Conference on (pp. 0387-0387). IEEE Computer Society.
[15] Giunchiglia, F., Mylopoulos, J., & Perini, A. (2003). The tropos software development methodology: processes, models and diagrams. In Agent-Oriented Software Engineering III (pp. 162-173). Springer Berlin Heidelberg.
[16] Gómez-Sanz, J. J., & Pavón, J. (2002). Agent Oriented Software Engineering with MESSAGE. Proceedings of the fourth international bi-conference workshop on agent-oriented information systems (AOIS 2002 at CaiSE'02) Toronto (Ontario, Canada), May 27-28.
[17] Iglesias, C. A., Garijo, M., González, J. C., & Velasco, J. R. (1996). A methodological proposal for multiagent systems development extending CommonKADS. In Proceedings of the 10th Banff knowledge acquisition for knowledge-based systems workshop (Vol. 1, pp. 25-1).
[18] Iglesias, C. A., Garijo, M., & González, J. C. (1999). A survey of agent-oriented methodologies. In Intelligent Agents V: Agents Theories, Architectures, and Languages: 5th International Workshop, ATAL'98 Paris, France, July 4–7, 1998 Proceedings (pp. 317-330). Springer Berlin Heidelberg.
[19] Jennings, N. R. (2001). An agent-based approach for building complex software systems. Communications of the ACM, 44(4), 35-41.
[20] Kellner, M. I., Madachy, R. J., & Raffo, D. M. (1999). Software process simulation modeling: why? what? how?. Journal of Systems and Software, 46(2), 91-105.
[21] Lehman, M. M. (1969). The programming process. Internal. IBM report.
[22] Lind, J. (1999). A process model for the design of multi-agent systems. Research report of German Research Center for AI (DFKI), number TM-99-03.
[23] Nwana, H. S., Ndumu, D. T., Lee, L. C., & Collis, J. C. (1999). ZEUS: a toolkit for building distributed multiagent systems. Applied Artificial Intelligence, 13(1), 129-185.
[24] Odell, J., Parunak, H. V. D., & Bauer, B. (2000). Extending UML for agents. Proc. of the Agent-Oriented Information Systems Workshop at the 17th National conference on Artificial Intelligence (AAAI), 2000.
[25] O'malley, S. A., & DeLoach, S. A. (2002). Determining when to use an agent-oriented software engineering In Proceedings of the Second International Workshop On Agent-Oriented Software Engineering (AOSE-2001), pages 188-205, Montreal, May 2001.
[26] Omicini, A. (2001). SODA: Societies and infrastructures in the analysis and design of agent-based systems. In Proceedings of the First International Workshop of Agent-Oriented Software Engineering, AOSE. Springer-Verlag, Berlin (pp. 185-193).
[27] Osterweil, L. (1987). Software processes are software too. In Proceedings of the 9th international conference on Software Engineering (pp. 2-13). IEEE Computer Society Press.
[28] Osterweil, L. J. (1997). Software Processes Are Software Too, Revisited: An Invited Talk on the Most Influential Paper of ICSE 9. In Proc. 19th Int'l Conf. Software Engineering (ICSE). (pp 540-548)
[29] Padgham, L., & Winikoff, M. (2003). Prometheus: A methodology for developing intelligent agents. In Agent-oriented software engineering III (pp. 174-185). Springer Berlin Heidelberg.
[30] Paulk, M. C., Weber, C., Curtis, B., & Chrisses, M. B. The Capability Maturity Model: Guidelines for Improving the Software Process. 1995. Addison Wesley.
[31] Perry D.E. and G.E. Kaiser. Models of software development environments. IEEE Transactions on Software Engineering, 17, March 1991.
[32] Sabas, A., Delisle, S., & Badri, M. (2002). A comparative analysis of multiagent system development methodologies: Towards a unified approach. In Third International Symposium From Agent Theory to Agent Implementation (AT2AI-3), Volume 2, pages 599–604.
[33] Shehory, O., & Sturm, A. (2001). Evaluation of modeling techniques for agent-based systems. In Proceedings of the Fifth International Conference on Au- tonomous Agents, pages 624-631. ACM Press.
[34] Sturm, A., & Shehory, O. (2004). A framework for evaluating agent-oriented methodologies. In Agent-Oriented Information Systems (pp. 94-109). Springer Berlin Heidelberg.

[35] Tveit, A. (2001). A survey of agent-oriented software engineering. In proc. Of NTNU Computer Science Graduate Student Conference, Norwegian University of Science and technology.
[36] Wood, M. F., & DeLoach, S. A. (2001). An overview of the multiagent systems engineering methodology. In Agent-Oriented Software Engineering (pp. 207-221). Springer Berlin Heidelberg.
[37] Wooldridge, M., Jennings, N. R., & Kinny, D. (2000). The Gaia methodology for agent-oriented analysis and design. Autonomous Agents and Multi-Agent Systems, 3(3), 285-312.
[38] Wooldridgey, M., & Ciancarini, P. (2001). Agent-oriented software engineering: The state of the art. In Agent-Oriented Software Engineering. Lecture Notes in AI Volume 1957. (pp. 1-28).
[39] Xu H. and S. M. Shatz, A Framework for Modeling Agent-Oriented Soft- ware, Proceedings of the 21st International Conference on Distributed Com- puting Systems (ICDCS 2001), April 16-19, 2001, Phoenix, Arizona, USA. IEEE Computer Society, pages 57-64, 2001.
[40] Xu, H. (2003). A Model-based approach for development of multi-agent software systems (Doctoral dissertation, University of Illinois).
[41] Zambonelli, F., Jennings, N. R., & Wooldridge, M. (2003). Developing multiagent systems: The Gaia methodology. ACM Transactions on Software Engineering and Methodology (TOSEM), 12(3), 317-370.